# Can the Higgs Boson Bootstrap itself ?

Bipin R. Desai and Alexander R. Vaucher

Department of Physics, University of California
Riverside, California 92521, USA

## Abstract

It is pointed out that Higgs bootstrap is inherent in the top-condensate models of Nambu and Bardeen et al in that the ladder sum of Higgs pole diagrams in one channel reproduces a Higgs pole in the crossed-channel. This result is exact whenever color flows simultaneously in both channel directions e.g. for $N_c = 1$. Bootstrap solutions for the Higgs boson mass are obtained that are compatible with the condensate models such as the top-color models which imply large top-Yukawa couplings at the top mass.

# I. Introduction

The suggestion [1], [2], [3] that the top-quark can couple to the Higgs boson more strongly than in the standard model leads to many interesting questions. One of these is the role the Higgs boson will play by providing a strong (attractive) force through t-channel exchange that can create bound states in $J^P = O^+$ channel (s-channel) of the top-antitop ($t\bar{t}$) scattering amplitude. A natural question to ask then is whether the Higgs boson will itself be among these bound states; that is, whether the dynamics will permit the Higgs boson to bootstrap itself.

Interestingly under certain circumstances a Higgs bootstrap is inherent in the $t\bar{t}$ condensate approach of Nambu [4] and Bardeen et al [5]. In the condensate model the Higgs boson is generated through a four-fermi interaction at a high scale by summing the $t\bar{t}$ bubble diagrams as in Fig. 1. This figure then represents the lowest order diagram for the Higgs-pole in the s-channel.

Consider now the t-channel Higgs exchange amplitude as indicated in fig.2. It is a t-channel sum of fermion bubbles in the same manner as the s-channel sum of fig.1. In terms of , $\lambda_t$, the top-Yukawa coupling and the Higgs mass, $m_H$, it can be expressed as follows

$$\frac{\lambda_t^2}{2} \frac{\bar{u}u\bar{v}v}{t - m_H^2} \tag{1}$$

where $u$ and $v$ are $t$ and $\bar{t}$ spinors respectively. If we now try to solve the bound state problem in the ladder approximation by iterating this amplitude then an interesting thing happens. In fig.3 each term in the ladder sum (the left column) is itself, according to fig.2, expressed as a sum of fermion bubbles (the right hand side). We notice that the first term in each sum on the right hand (i.e. the first column on the right after the equal-sign) when added up will give the Higgs amplitude <u>in the s-channel as in fig.1</u>, the subsequent terms on the right being unitarity corrections. That is, a sum of t-channel Higgs-exchanges produces an s-channel Higgs amplitude (see fig.4).

This implies that the Higgs exchange potential will produce Higgs boson as a bound state in the s-channel provided the coupling is strong enough. Therefore a Higgs bootstrap is inherent in the four fermi condensate model, at least in the ladder approximation. Of course, since color flows either in the s-channel direction or the t-channel but not both simultaneously in these diagrams, a strict bootstrap occurs only when $N_c = 1$.

We note, however, that in the condensate model once the gap equation is imposed, the expression for the mass of the Higgs boson itself in terms of top-mass i.e.

$$m_H = 2m \qquad (2)$$

is independent of $N_c$. The coupling, of course, depends on $N_c$ [5].

How strong the top-Yukawa coupling should be to create bound states and therefore to accomplish the bootstrap will be considered below within the framework of the ladder approximation.

Instead of carrying out a general bootstrap program by varying $m_H$ and $\lambda_t$, we will consider only the question whether the Higgs boson mass (2) reproduces itself.

## II. The Bound State Problem

The bound state problem in the ladder approximation has been studied in certain field theories [6], in non-relativistic potential theories with Yukawa potentials [7] and in S-matrix theories [8], [9]. The solution is obtained through dispersion relations by imposing the known analytic and unitarity properties of partial wave scattering amplitudes expressed in a formalism of the ratio N/D. The zeroes of D, then determine the bound states.

Ghergetta [10] has shown that the dispersion relations methods of the type indicated above are better suited than the traditional methods in Nambu Jona-Lasinio type models [11], specifically in the top quark condensation models of Nambu [4] and Bardeen et al [5].

For our bound state calculation we use the Jacob and Wick [12] helicity formalism for states, $|\lambda_1, \lambda_2\rangle$, with helicities $\lambda_1 = \pm\frac{1}{2}$, $\lambda_2 = \pm\frac{1}{2}$, with the scattering amplitude $ab \to cd$ given by

$$T_{cd,ab}(E) = \frac{1}{p}\sum_j \left(j+\frac{1}{2}\right)\langle\lambda_c\lambda_d|T_j(E)|\lambda_a\lambda_b\rangle e^{i(\lambda-\mu)\phi}d^j_{\lambda\mu}(\theta)$$

where $\lambda = \lambda_a - \lambda_b$, $\mu = \lambda_c - \lambda_d$. For $J^P = O^+$ in the s-channel, the appropriate linear combination of the states is

$$|+\rangle = \frac{|++\rangle + |--\rangle}{\sqrt{2}}$$

where + and − indicate $+\frac{1}{2}$ and $-\frac{1}{2}$ respectively. For our scattering amplitude, therefore, $\lambda = 0 = \mu$ and

$$d^j_{\lambda\mu} = P_j(\cos\theta)$$

Denoting this amplitude as $T_+$, we write

$$T_+(E,\theta) = \frac{1}{p}\sum_j \left(j+\frac{1}{2}\right)\langle+|T_{+j}(E)|+\rangle P_j(\cos\theta) \tag{3}$$

The Higgs-exchange amplitude $T_+^H$ can be obtained from (1) as

$$T_+^{(H)}(E,\theta) = -\lambda_t^2 \frac{\left[m^2(1+\cos\theta) - E^2(1-\cos\theta)\right]}{t - m_H^2} \tag{4}$$

where $t = -2p^2(1-\cos\theta)$ is the momentum transfer and $p$ and $\theta$ the center of mass momentum and scattering angle respectively.

The partial-wave projection is given by

$$T_{+j}^{(H)}(E) = \frac{1}{2}\int_{-1}^{1} d\cos\theta\, T_{+}^{(H)}(E,\theta)\, P_j(\cos\theta) \tag{5}$$

If we write $m_H = 2m$ (condensate value), and

$$x = \frac{p^2}{m^2} \tag{6}$$

then the $j = 0$ value is given by

$$T_{+0}^{(H)} = \frac{\lambda_t^2}{2}\left[\left(1+\frac{2}{x}\right)-\left(\frac{2}{x}+\frac{2}{x^2}\right)\ln(1+x)\right] \tag{7}$$

where we have assumed $\lambda_t^2$ to be a constant (we will return to this assumption later).

In the following we will attempt to solve our Higgs problem through the N/D method mentioned earlier. Typically, in this formalism, a partial wave amplitude

$$T_l = \frac{e^{i\delta}\sin\delta}{\rho} \tag{8}$$

$$= \frac{1}{\rho\cot\delta - i\rho}$$

where $\delta$ is the phase-shift and $\rho$ the phase space factor, is expressed as a ratio

$$T_l = \frac{N}{D} \tag{9}$$

where $D$, has a branch cut from the threshold at $p^2 = 0$ to $\infty$ (right hand cut) as it acquires a phase given by (5). But it is real for $p^2 < 0$. $N$ is real for $p^2 > 0$ but has a (left hand) cut arising from singularities of the partial wave projection of the t-channel exchange amplitude. This cut extends from $-\infty$ to $v_0$, where $v_0$ is related to the mass of the exchanged particle ($v_0 = -\frac{m_H^2}{4}$ in our case). In terms of $x$ defined in (6) the right hand cut is along $(0, \infty)$. And the left hand cut along $(-\infty, -1)$ with $m_H = 2m$, as can be confirmed from (7).

The discontinuity of $D$ across the right-hand cut is easily seen from (8) and (9) to be $-\rho N$. The left hand discontinuity of $N$ is $D \operatorname{Im} T_l$, where "Im" indicates imaginary part in the interval $(-\infty, v_0)$.

Specifically, normalizing the amplitude by taking $D \to 1$ as $x \to \infty$, one writes

$$D(x) = 1 - \frac{1}{\pi} \int_0^\infty \frac{dx'}{x' - x} \rho\, N(x') \qquad (10)$$

$$N(x) = \frac{1}{\pi} \int_{-\infty}^{-1} \frac{dx'}{x' - x} D(x')\, \mathrm{Im}\, T_l(x') \qquad (11)$$

with
$$\rho = \frac{1}{16\pi} \sqrt{\frac{x}{x+1}}$$

$N$ is usually taken simply as the partial wave projection of the t-channel one particle exchange amplitude (in our case, the partial wave projection of the amplitude (1) given by (7)) without going through the dispersion integrals in (11). This is what we will also assume for the Higgs exchange [9]. Therefore, we will take

$$N(x) = T^H_{+0}(x).$$

We note from (7) that $N(x)$ has the proper analytic structure. We need now to obtain $D(x)$. Before we insert $N(x)$ given above in the dispersion integral (10) we note that $\lambda_t$, which was assumed constant, actually evolves according to the renormalization group equations as

$$\mu^2 \frac{d\lambda_t}{d\mu^2} = b\lambda_t^3 \qquad (12)$$

where the top-Yukawa coupling $\lambda_t$ is evaluated at mass $\mu$. Since $\lambda_t$ is no longer restricted to its standard model value ($\approx 1$) at the top mass but can be much larger, according to the top-color models [1],[2],[3], the above equation implies a rapid increase in $\lambda_t$ for larger values of $\mu^2$ [13].

In (4) and (12) we note that since $\lambda_t = \lambda_t(\mu^2)$, where $\mu^2 = t$, the t-dependence of $\lambda_t$ in the Higgs-exchange amplitude given by (4) must be taken into account in the partial wave projection (5), and, therefore, in (7). The integration in (5) of $\cos\theta$ from $-1$ to $+1$ converts to

$$\frac{1}{2}\int_{-1}^{1} d\cos\theta = \frac{1}{4p^2}\int_{-4p^2}^{0} dt$$

This implies that it is the region $\mu^2 < 0$ which is relevent for $\lambda_t(\mu^2)$. From the renormalization group equations (12) we note that $\lambda_t(\mu^2)$ does not depend on the sign of $\mu^2$ and its explicit $\mu^2$ dependence can be determined from the appropriate boundary conditions.

Clearly then, depending on the functional form of $\lambda_t(\mu^2)$, the above integration and the subsequent dispersion relations integral (10) for $p^2$ from 0 to $\infty$ ( i.e. $x$ from 0 to $\infty$) from can be very complicated. A reasonable approximation would be to replace $\lambda_t^2$ by some sort of average, $\langle \lambda_t^2 \rangle$, over $\mu^2$.
Thus we change $\lambda_t^2 \to \langle \lambda_t^2 \rangle$.

It is also important to note that in addition to the possibility that $\lambda_t$ can be large at the top mass $\mu = m$ [1], [2], [3], its evolution given by (12), can make $\lambda_t$ even larger as $\mu^2$ increases. Bardeen et al [5] have emphasized that if the $t\bar{t}$ system forms a bound state, then $\lambda_t$ will blow up as $\mu \to \Lambda$ where $\Lambda$ is the appropriate composite scale. Therefore,

$$\langle \lambda_t^2 \rangle = \text{very large}.$$

We will return later to the question of determining $\langle \lambda_t^2 \rangle$.

From (7) we note that

$$T_{+0}^{(H)} \to \text{constant} \left(= \frac{\lambda_t^2}{2}\right) \text{ as } x \to \infty$$

The integral in (10) for D, therefore, will not converge making it necessary to introduce a subtraction to represent the unknown high scale dynamics with D normalized so that D=1 at the subtraction point. Since we are looking for possible bound states i.e. zeroes in D in the region $-1 < x < 0$ it is essential that the subtraction point not include this region. We will take, in the following, the subtraction point at $x = -x_1$, such that $x_1 \geq 1$, and discuss the bound state energies as well as any sensitivities of the results to the subtraction point.

Taking account of the above comments we express D given by (10), using (7) as

$$D(x) = 1 - \frac{\langle \lambda_t^2 \rangle}{32\pi^2}(x+x_1)\int_0^\infty \frac{dx'}{(x'-x)(x'+x_1)\sqrt{x'(x'+1)}}\left[(x'+2) - \frac{2(1+x')\ln(1+x')}{x'}\right] \quad (13)$$

We write the above relation as,

$$D(x) = 1 - \frac{\langle \lambda_t^2 \rangle}{32\pi^2}(x+x_1)\, I(x,x_1) \quad (14)$$

where $I(x,x_1)$ is the integral on the right hand side of expression (13). To accomplish a bootstrap vis-à-vis the condensate model with $m_H = 2m$ (i.e. $x = 0$) we must have

$$D(0) = 0$$

Therefore the Yukawa coupling must satisfy

$$\frac{\langle \lambda_t^2 \rangle}{16\pi^2} = \frac{2}{x_1 I(0, x_1)} \tag{15}$$

A plot of the right side of equation (15) is given in fig. 5 for $x_1 > 1$. We find that it is insensitive to the precise choice of $x_1$ for $x_1 > 2$, being roughly a constant over a large range of $x_1$, decreasing slowly, logarithmically. For a typical $x_1$ associated with an s-channel scale of $\Lambda \approx 1$ Tev ($x_1 = \frac{\Lambda^2}{4m^2}$), for which $x_1 \approx 8$, we obtain

$$\frac{\langle \lambda_t^2 \rangle}{16\pi^2} \approx 1.5 \tag{16}$$

Thus a Higgs bootstrap is possible as long as the averaged coupling constant is consistent with (16). Next we consider whether this value is reasonable based on the renormalization group equations.

## III. Estimating $\langle \lambda_t^2 \rangle$

To estimate $\langle \lambda_t^2 \rangle$ we note that the solution for (12) with the compositeness boundary condition $\lambda_t \to \infty$ as $\mu \to \Lambda$ is given by [5]

$$\lambda_t^2(\mu) = \frac{1}{2b \ln\left(\frac{\Lambda^2}{\mu^2}\right)} \quad , b > 0 \tag{17}$$

If we try to evaluate $\langle \lambda_t^2 \rangle$ by defining it as a simple average,

$$\langle \lambda_t^2 \rangle = \frac{1}{\Lambda^2} \int_{\mu_0^2}^{\Lambda^2} d\mu^2 \, \lambda_t^2(\mu^2) \tag{18}$$

then we find that substituting (17) in (18) will make the integral infinite because of the singularity at $\mu = \Lambda$ in (17).

A more convergent way to estimate $\langle \lambda_t^2 \rangle$ may be by extrapolating $\lambda_t^2$ at $\mu^2 = \mu_0^2$ linearly through $\mu^2 = \Lambda^2$ by using (12) and then taking the average as given by (18). If we write $\lambda_t^2(\mu_0^2) = \lambda_0^2$ we obtain from (12)

$$\lambda_t^2(\mu) = \lambda_0^2 + 2b\lambda_0^4 \ln\left(\frac{\mu^2}{\mu_0^2}\right) \tag{19}$$

Inserting the above expression in (18) we obtain, for large $\Lambda$,

$$\langle \lambda_t^2 \rangle \approx \lambda_0^2 + 2b\lambda_0^4 \ln\left(\frac{\Lambda^2}{\mu_0^2}\right) \tag{20}$$

For our Higgs problem, we take $\mu_0 = m$, the top mass, then $\lambda_0$ is the top-Yukawa coupling at the top-mass ($\lambda_0 = \lambda_t(m^2)$).

For the condensate model, Pagles and Stockar [14], and Gherghetta [10], have derived a relation for the coupling in terms of the composite scale $\Lambda$

$$\lambda_0^2 = \lambda_t^2(m^2) = \frac{16\pi^2}{N_c} \frac{1}{\ln\left(\frac{\Lambda^2}{m^2}\right)} \tag{21}$$

Eliminating the logarithm from (20), using (21) (with $N_c = 1$) we obtain, using the standard model value for $b$, the following estimate for the top-Yukawa coupling at the top mass,

$$\lambda_0 \approx 4.87 \tag{27}$$

This is a rather large value but consistent with the top-color model estimates [1], [2], [3].

We, therefore, conclude that the Higgs bootstrap which is inherent in the ladder diagram approach of the condensate models is actually feasible for sensible values of $\lambda_0$.

We thank Professor Jose Wudka for his helpful comments. This work was supported in part by the U.S. Department of Energy under Contract No: DE-F603-94ER40837.

# References

[1] C. T. Hill, Phys. Lett. B 266, 419 (1991)

[2] C. T. Hill, Phys. Lett. B 345, 483 (1995); D. Kominis Phys. Lett. B 358, 312 (1995); J.D. Wells, hep-ph/9612292; M. Spira and J.D. Wells, hep-ph/9711410.

[3] R.S. Chivkula, B.A. Dobrescu, H. Georgi and C.T. Hill, hep-ph/9809470.

[4] Y. Nambu, report EFI 88-39 (July1988), published in the proceedings of the *Kazimierz 1988 Conference on New Theories in physics*, ed. T. Eguchi and K. Nishijima; in the proceedings of the *1988 International workshop on New Trends in Strong Coupling Gauge Theories, Nagoya, Japan*, ed. Bando, Muta, and Yamawaki (World Scientific, 1989); report EFI-89-08 (1989); Also see V.A. Miransky, M. Tanabashi and K. Yamawaki, Mod. Phys. Lett. A4 (1989) 1043; Phys. Lett. B 221 (1989) 177; W.J. Marciano, Phys. Lett. 62 (1989) 2793.

[5] W.A. Bardeen, C.T. Hill and M. Lindner, Phys. Rev. D41, 1647 (1990).

[6] B. Lee and R. Sawyer, Phys. Rev. 127, 2266 (1962); Also see D. Amati, S. Fubini and A. Stanghellini, Phys. Lett. 29 (1962).

[7] R. Blankenbecler and M.L. Goldberger, Phys. Rev. 126, 766 (1962); M.L. Godberger, M.T. Grisaru, S.W. McDowell and D. Wong, Phys. Rev. 120, 2250 (1960). Also see R.G. Newton.

[8] G.F. Chew, *S-matrix Theory of Strong Interactions*, Benjamin, New York (1962).

[9] S. Gasiorowitz, *Elementary Particle Physics*, John Wiley & Sons Inc, New York (1966).

[10] T. Gherghetta, Phys. Rev. D50, 5985 (1994).

[11] Y. Nambu and G. Jona-Lasinio, Phys. Rev. 122, 345 (1961); ibid Phys. Rev 124, 246 (1961).

[12] M.Jacob and G.C. Wick, Ann. Phys. (NY) 7, 404 (1959).

[13] We are taking the conventional definition, $\lambda_t = \frac{\sqrt{2}\, m}{v}$, for the top-Yukawa coupling so that for the standard model value, $v = 246$ Gev, we have $\lambda_t \approx 1$ at the top mass. If, on the other hand, $v$ is much smaller than this value, as in the top-color models, then $\lambda_t$ will be quite large.

[14] H. Pagels and S. Stockar, Phys. Rev. D20, 2947 (1979).

Figure Captions:

Fig. 1.  Higgs boson as a fermion bubble summation in the s-channel.

Fig. 2.  Higgs boson as a fermion bubble summation in the t-channel.

Fig. 3.  Individual terms in the ladder summation of t-channel Higgs-exchange diagrams expressed in terms of fermion bubble diagrams.

Fig. 4.  Graphical representation of the bootstrap result.

Fig. 5.  $\dfrac{2}{x_1 I(0, x_1)} \left( = \dfrac{\langle \lambda_t^2 \rangle}{16\pi^2} \right)$ vs. $x_1$.

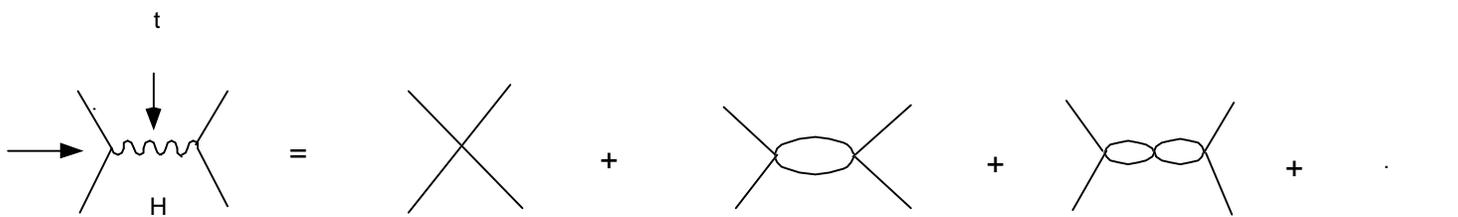

fig. 1

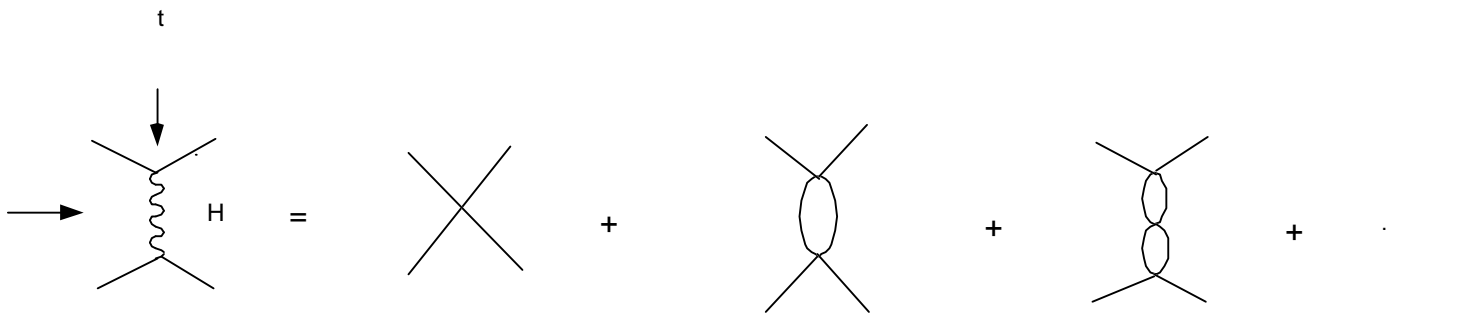

fig. 2

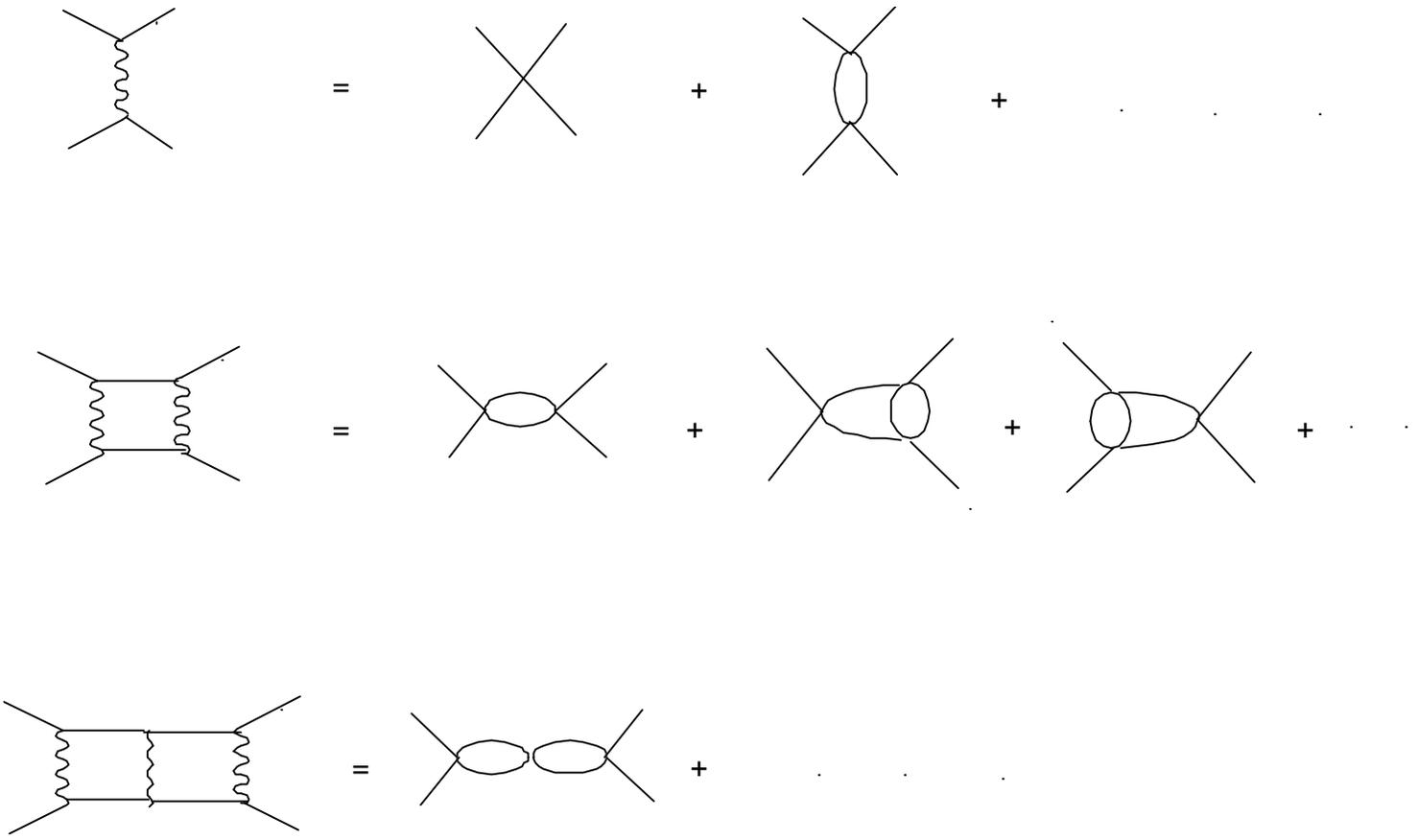

Fig. 3.

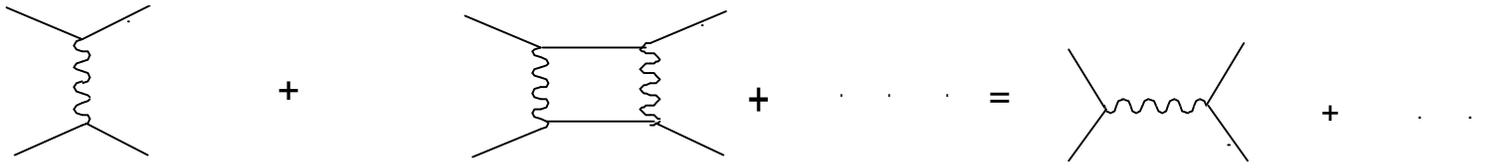

Fig. 4

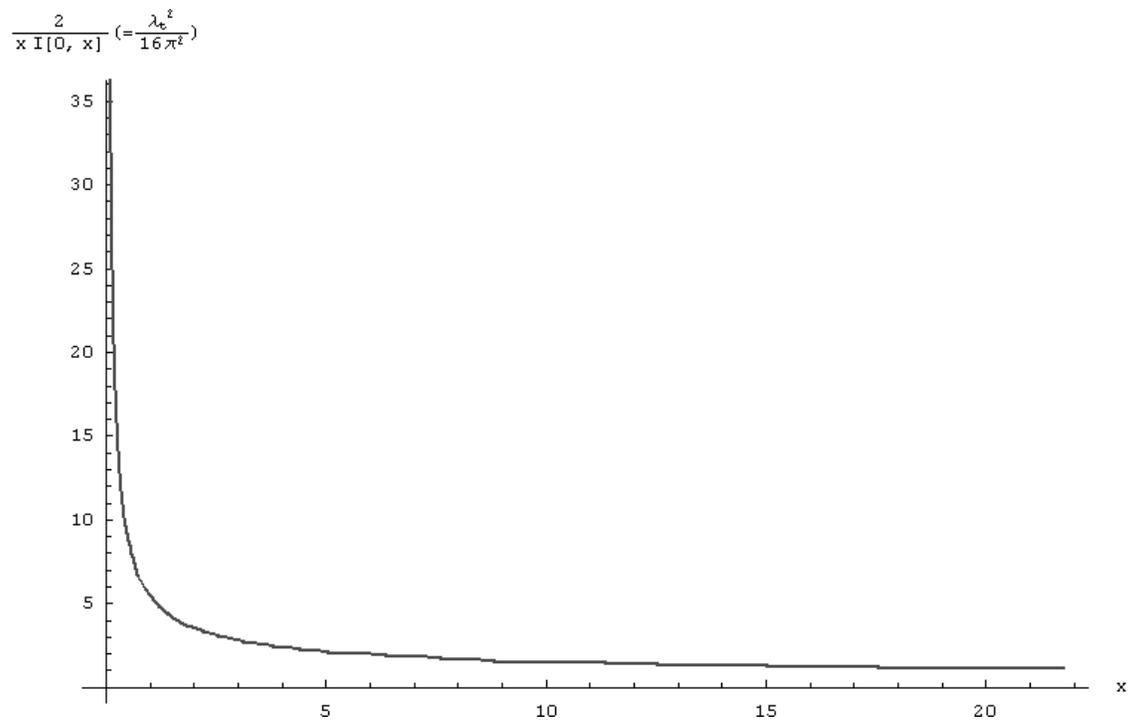

Fig. 5